\documentclass[useAMS,usenatbib]{mn2e}
\usepackage{graphicx}

\def\ergs{${\rm erg\,cm^{-2}\,s^{-1}}$ }
\def\ergse{${\rm erg\,cm^{-2}\,s^{-1}}$}
\def\ucr{${\rm cts\,s^{-1}}$ }
\def\ucre{${\rm cts\,s^{-1}}$}
\def\ulum{${\rm erg\,s^{-1}}$ }
\def\ulume{${\rm erg\,s^{-1}}$}

\def\unhe{${\rm cm^{-2}}$}

\def\qso{GB\,1508+5714 }

\def\ucpx{cts\,pixel$^{-1}$ }
\def\ucpxe{cts\,pixel$^{-1}$}

\title[Extended X-ray emission in GB\,1508+5714 at $z=4.3$]
{Extended X-ray emission 
in the high redshift quasar GB\,1508+5714 at $z=4.3$}
\author[W. Yuan]{W. Yuan$^{1}$\thanks{E-mail:
wmy@ast.cam.ac.uk (wy)},
A.C. Fabian$^{1}$,
A. Celotti$^{2}$
and
P.G. Jonker$^{1}$\\
$^{1}$University of Cambridge, Institute of Astronomy, 
Madingley Road, Cambridge, CB3 0HA\\
$^{2}$SISSA, via Beirut, 2-4, 34014 Trieste, Italy}

\begin{document}

\date{Accepted ??? Received ??? in original form Aug. 15, 2003}

\pagerange{\pageref{firstpage}--\pageref{lastpage}} \pubyear{2003}

\maketitle

\label{firstpage}

\begin{abstract}
We report the discovery of extended X-ray emission around the powerful
high-redshift quasar \qso at $z$=4.3, revealed in a long Chandra ACIS
observation. The emission feature is 3--4\,arcsec away from the 
quasar core, which corresponds to a projected distance of about 25\,kpc. 
The X-ray spectrum is best fitted with a power law of photon index
$1.92\pm0.35$ (90 per cent c.l.).
The X-ray flux and luminosity reach
$9.2\,10^{-15}$\,\ergs (0.5--8\,keV) and
$1.6\,10^{45}$\,\ulum (2.7--42.4\,keV rest frame, 
$\Omega_{\Lambda}$=0.73, $\Omega_{\rm m}$=0.27, 
$H_0$=71\,km\,s$^{-1}$\,Mpc$^{-1}$),
which is about 2 percent of the total X-ray emission of the quasar.
We interpret the X-ray emission as inverse Compton scattering of Cosmic
Microwave Background photons. 
The scattering relativistic electron population could 
either be a quasi-static diffuse cloud fed by the jet, 
or an outer extension of the jet with a high bulk Lorentz factor.
We argue that the lack of an obvious detection of radio emission
from the extended component could be a consequence of 
Compton losses on the electron population, or of a low magnetic field.
Extended X-ray emission produced by inverse Compton scattering may be
common around high redshift radio galaxies and quasars, 
demonstrating that significant power is injected into their
surrounding by powerful jets.
\end{abstract}


\section{Introduction}

High redshift active galactic nuclei (AGN) are powerful tools with which to 
study the evolution of super-massive black holes (SMBH) and their host
galaxies at an early epoch.
One important issue concerns the feedback from SMBH to the host
galaxy by means of kinetic energy input carried in energetic particle flows
in the form of jets.
Relativistic jets are inferred to be present in high-redshift 
radio-loud AGN, 
particularly in blazar-like objects (i.e.\ with a jet oriented close to
the line of sight), from their extremely high luminosities
($\sim 10^{47}$\,\ulume)
and variability in both the X-ray and radio bands 
(e.g.\ Fabian et al.\ 1997, 1999).

In the X-ray band, thanks to the sub-arcsec spatial resolution of 
the Chandra X-ray Observatory, 
direct evidence of resolved X-ray jets and extended emission 
has been obtained in dozens of radio galaxies and quasars
(e.g.\ Chartas et al.\ 2000, Schwartz et al.\ 2000, 
Worrall et al.\ 2001,
Fabian, Celotti \& Johnstone 2003b, Harris et al.\ 2002, 
Siemiginowska et al.\ 2003).
Models have been proposed to explain the X-ray emission
on such scales involving synchrotron, 
inverse Compton scattering of photons of
the synchrotron, Cosmic Microwave Background (CMB) and  
radiation from the AGN nucleus
(e.g.\ Celotti et al.\ 2001, Tavecchio et al.\ 2000,
Harris \& Krawczynski 2002).
Inverse Compton scattering of the CMB is particularly
relevant at high redshifts since, as pointed out by Schwartz (2002a),
the X-ray surface brightness of jets is roughly constant with redshift 
as a consequence of the steep $(1+z)^4$ increase 
in the energy density of the CMB.
Another factor to consider at high redshifts (above $\sim 2$) is that
the angular size actually increases with redshift.
While there have been efforts to search for X-ray jets at high redshifts
e.g.\ up to $z\sim6$ (Schwartz 2002b), 
the highest redshift at which a convincing detection 
has been obtained so far 
is $z$=3.8 for the radio galaxy 4C\,41.17 as reported
by Scharf et al.\ (2003) recently.

The quasar \qso at $z$=4.301 (Hook et al.\ 1995) is one of a dozen
quasars at redshifts above 4 with strong radio emission
($f_{\rm 5GHz} > 10{\rm mJy}$). 
Variability in the X-ray and radio bands, 
as well as a high X-ray luminosity of $\sim 10^{47}$\,\ulum
(Moran \& Helfand 1997) suggests
relativistically beamed X-ray and radio emission of a blazar type.
In several high redshift quasars soft X-ray spectral flattening
has been found 
(e.g.\ Boller et al.\ 2000, Yuan et al.\ 2000, Fabian et al.\ 2001).
However, in our XMM observation of \qso (paper in preparation) 
no evidence for such an effect was found (see also Moran \& Helfand 1997).

We report here on the discovery of extended X-ray emission in \qso
  from a long Chandra ACIS-S observation (Sect.\,2) 
and discuss its possible origin (Sect.\,4). 
We used $H_0$=71\,km\,s$^{-1}$\,Mpc$^{-1}$,
$\Omega_{\Lambda}$=0.73, and $\Omega_{\rm m}$=0.27
to calculate luminosities and distances throughout the paper.
Errors quoted are at the $1\,\sigma$ level unless stated otherwise.

\section[]{Observations and data analysis}
The quasar \qso was observed with Chandra with its back-illuminated
CCD ACIS-S3 in the faint mode for a duration of 91.8\,ksec
starting from 2001 June 10.
The data were taken from the Chandra archive 
(PI: Paerels, Paerels et al.\ 2002).  
The target was placed close to the nominal aim-point with a small offset of
20\,arcsec.  The data analysis was performed using the standard CIAO tools
(version 2.3, caldb version 2.21).  The good exposure time is
88.97\,ksec after filtering the data.  
A level 2 events file was created from level 1 data
following the standard procedure.
Events with grade of 0,2,3,4,6 were selected.
Only events with energy in the 0.3--8\,keV range were included.

\subsection[]{Extended X-ray emission feature}
\begin{figure}
   \includegraphics[angle=0,width=\hsize]{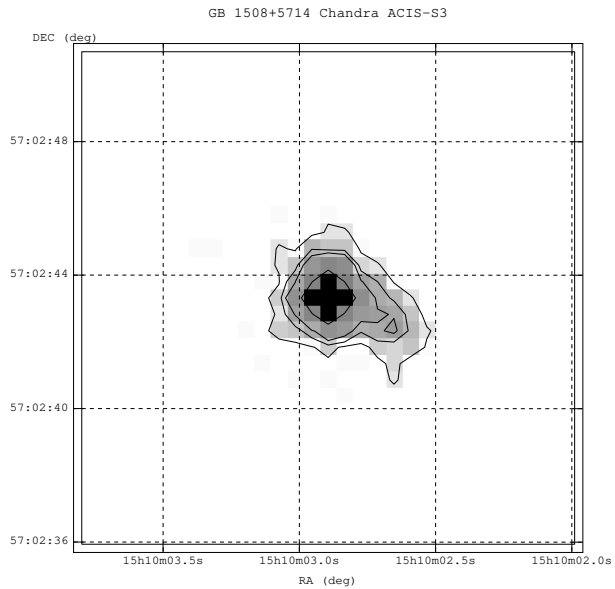} 
      \caption{\label{fig:ximg_cnt}
      X-ray image of the quasar \qso in the 0.3--2\,keV band on
      15$\times$15\,arcsec scales.
      The bin size is 0.5\,arcsec, i.e.\ the original CCD pixel
      is used  (logarithmic grey scale, cutoff range 2--200\,\ucpxe).
      The X-ray emission is extended in the south-west
      direction up to 4\,arcsec (a projected distance of 28\,kpc).     
      The contours are at levels of 2.5, 10, 22, 180\,\ucpxe.
      } 
   \end{figure}

The X-ray image of \qso in the 0.3--2\,keV band is shown in
Fig.\,\ref{fig:ximg_cnt},
with the bin size as 1 original pixel of the CCD, i.e.\ 0.5\,arcsec.
We used the 0.3--2\,keV band for spatial analysis as the Chandra 
point spread function (PSF) is best understood within this energy range.
The core of the X-ray source is unresolved 
(the source profile being consistent with that of the PSF) 
and coincident with the optical/radio position of the quasar
(15h10m02.9s +57d02m43s J2000).
In addition, to the south-west of the core the X-ray emission is
apparently resolved and extended.
This extended part of the emission is peaked at 
15h10m02.6s +57d02m42s 
(indicated by a smaller contour of 22\,\ucpx level 
in Fig.\,\ref{fig:ximg_cnt}),
which is $\sim 3$\,arcsec away from the 
position of the quasar core and has a position angle of  $250^{\circ}$ 
relative to the core.
At a redshift of 4.3, this angular distance 
corresponds to a projected distance of 21\,kpc.
The feature extends up to $\sim 4$\,arcsec (28\,kpc) from the core 
along the direction defined by this position angle, 
and spans $\sim 2$\,arcsec (14\,kpc) in the transverse direction.

\begin{figure}
   \includegraphics[angle=0,width=\hsize]{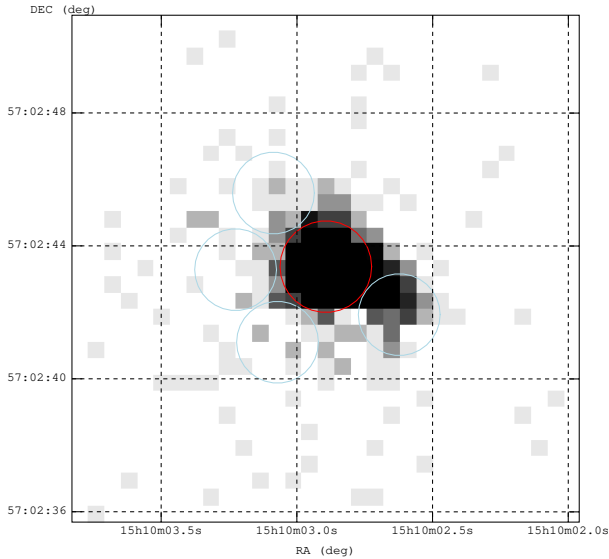} 
      \caption{\label{fig:extract}
      The same image as in Fig\,\ref{fig:ximg_cnt} with a different
      grey scale cutoff range (0--80\,\ucpxe).
      Over-plotted is the circle of 95 per cent encircled energy for the
      quasar core (central, 1.4\,arcsec in radius).
      Also plotted are
      the circular regions used for extracting the 
      extended emission (west-most) and the background (eastern three).
      } 
   \end{figure}

The source counts of the extended emission were extracted from a circle
with a radius of 2.5\,pixel (1.25\,arcsec), 
as shown in Fig.\,\ref{fig:extract}.
It covers the extended emission region and lies outside the
circle of 95 per cent encircled energy of the central point source with
a radius of 1.4\,arcsec (the circle around the core in 
 Fig.\,\ref{fig:extract}).
To extract background counts, we have to take into account 
the contamination of photons from the nearby core.
We define three circles as background regions
which have the same sizes and are located
at the same radial distances from the quasar core 
as the source region (see Fig.\,\ref{fig:extract}).
Since the PSF of the
point source is nearly symmetric (the object is almost on--axis), 
the contamination from the quasar core is expected to be eliminated to a
large extent with this background subtraction.
In the 0.3--8\,keV band,
the total number of counts in the extended source region is 158. 
The averaged number of counts 
in each of the three background regions is 22.3,
consistent with what is expected from the contribution of the
core for the Chandra PSF.
The net source counts of the extended emission 
(with possible contamination from the quasar core excluded)
is estimated to be 136$\pm$13, 
yielding a highly significant detection (11\,$\sigma$).
Its count rate is 1.50$\pm$0.14$\times10^{-3}$\,\ucr (0.3--8\,keV),
which is $\sim$2.5 per cent of that of the quasar (0.06\,\ucre).

We do not consider this extended emission feature as an artifact.
In the data validation and verification report
the observation and data processing were noted as having 
a good aspect solution and no identified problem.
Furthermore, the count rates of the extended emission show no sign
of time variability on any timescale during the observation.

We attempted to distinguish whether the extended emission is 
point-like or diffuse by modelling the 2-D quasar core emission 
using the Chandra PSF and then subtracting the model from the image data.
However, the result is inconclusive, given
the small number of net source counts of the extended
emission in contrast to the nearby bright quasar core,
and the uncertainty in modelling the PSF.
Future deep X-ray observations with higher spatial resolution 
are needed to resolve this emission.

\subsection[]{X-ray spectrum of the extended emission}
\begin{figure}
   \includegraphics[angle=270,width=\hsize]{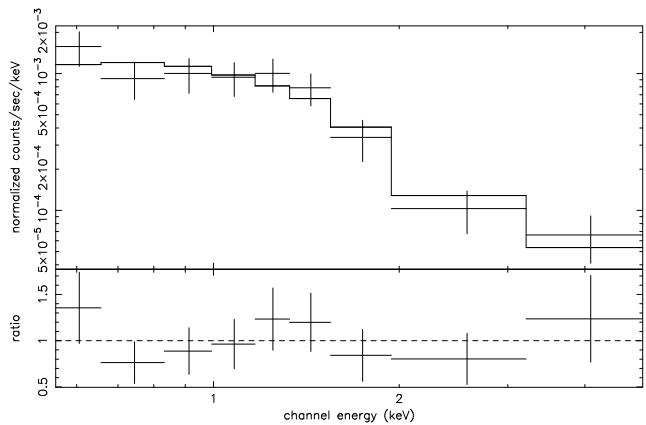}
      \caption{\label{fig:xspec}
      The X-ray spectrum of the extended emission feature and
      the best-fit absorbed power law model (upper panel).
      The ratio of the data to the model is also shown (lower panel).
      }
   \end{figure}

The spectra of the extended emission and the background
were extracted from the regions described above.
The RMF and ARF files were created at the source position. 
We performed spectral fitting using XSPEC (v.11.2).
The effect of degradation in the ACIS quantum efficiency was
taken into account by applying the {\it apply\_acisabs} provided 
in CIAO (Chartas \& Getman 2003).
Given the small number of source counts, 
the spectrum was binned to have at least 15\,counts per energy bin,
and the C-statistic is adopted in the spectral fitting.
Absorption with Galactic column density (1.6$\times10^{20}$\,\unhe) 
was included and the value was fixed in the fit.
The background-subtracted spectrum is best fitted with
a power law model, (C-stat=4.2 for 9 bins),
yielding a photon index of
$1.92_{-0.33}^{+0.38}$ (90 per cent confidence level). 
The spectrum and fitted residuals are shown in Fig.\ref{fig:xspec}.
A redshifted thermal plasma model (Raymond-Smith) 
gave a slightly worse but acceptable fit (C-stat=5.2 for 9 bins)
with a rest frame temperature $T=17.2^{+21.1}_{-7.4}$\,keV.

The unabsorbed flux density in the 0.5--8keV band is 
$9.2\times10^{-15}$\,\ergse, using the best-fit power law model,
which corresponds to a 2.7--42.4\,keV luminosity of 
$1.6\times10^{45}$\,\ulum in the quasar rest frame.

\section[]{Constraints from other wavebands}
In the radio band, the quasar was detected as a radio source at
low and high frequencies, e.g.\ 222\,mJy at 365\,MHz (Douglas et al.\ 1996),
279\,mJy at 4.85\,GHz (Becker et al.\ 1991).
The radio emission \qso was unresolved 
with VLA A-configuration at 1.4 (234\,mJy) and 8.4\,GHz (152\,mJy) 
with 5-min observations (Moran \& Helfand 1997).
On mas scales the quasar remains  unresolved with VLBI observations
(the EVN, Frey et al.\ 1997).
The compactness of the radio emission suggests that the postulated radio 
jet is closely aligned to the line of sight.

The optical imaging data for \qso taken with 
the Wide-Field-Camera at the Isaac Newton Telescope 
show no emission at the position of the extended X-ray emission
at $5\,\sigma$ limiting magnitudes of
23.20 and 23.57 in the $i'$ and $g'$-band, respectively.
These data also suggest that 
it is unlikely that the extended X-ray emission is coming from 
a foreground or background AGN.
This is because at this X-ray flux level
the majority of AGNs have I-band magnitudes brighter than $I\sim$23.3
(corresponds to $f_{\rm x}$/$f_{\rm opt}\leq$10), 
as found in the Chandra deep surveys (e.g.\ Alexander et al.\ 2001).
In the following section of this paper
we consider only the most likely scenario 
that the extended X-ray emission is associated with the quasar.

\section{The emission mechanism and nature of the extended emission}

We have found a blob of extended X-ray emission about 3 arcsec away
from the powerful blazar GB\,1508+5714. This corresponds to a projected
distance of about 21\,kpc, but could be 100\,kpc or more if it is the
outer part of the blazar jet. Interpretation of this X-ray emission is
not straightforward given the paucity of information from other wavebands. 

The most obvious emission mechanism is inverse Compton scattering of
Cosmic Microwave Background photons. At the quasar redshift of
4.3 the energy density in the CMB is 790 times that now. The
relativistic electron population scattering the CMB photons is
plausibly due to the blazar jet. It can either be a quasi-static
diffuse cloud fed by the jet, or an outer extension of the jet with a
bulk Lorentz factor $\Gamma\sim 10$. This last possibility has been
proposed for the X-ray jets seen around several lower redshift quasars
(e.g.\ Celotti et al.\ 2001, Tavecchio et al.\ 2000,
Siemiginowska et al.\ 2003).

Infrared emission from the blazar itself may contribute to 
the seed photons for inverse Compton scattering 
(e.g.\ Brunetti et al.\ 1997) but the emission would have
to be very luminous ($\sim 10^{48}$\,\ulume)
to compete with the CMB at that redshift. 
We have found no published far infrared measurements 
of this object with which to make an estimate
of any such contribution.

An important issue is to explain why the X-ray blob is not seen at
radio wavelengths. High resolution radio imaging by the EVN (Frey et
al.\ 1997) shows that the source is unresolved on the 10 milliarcsec scale
at 5\,GHz. The flux in that component is within 2 mJy of the total flux
seen with the VLA. Although the source may vary between the
observations, we shall assume that the radio emission from this
component is less than a few mJy. 

The lack of a detection of the extended component could be a
consequence of Compton losses on the electron population, or of a low
magnetic field. If the electrons have no relativistic bulk motion,
then the X-ray emission requires electrons with Lorentz factor
$\gamma\sim 10^3\gamma_3$ with $\gamma_3\sim 1$. The magnetic field in
the same region required to produce 5~GHz radio emission in our rest
frame is $B\sim 10^{-2}\gamma_3^{-2}$~G. The energy density of this
magnetic field is about $10^4$ times that in the CMB, which would be
reflected in the radio luminosity being $10^4$ times that in the
X-ray, rather than the observed limit of $0.1$. Consequently the
magnetic field must be much lower and $\gamma$ much higher in order to
produce 5\,GHz radio emission.

The lifetime of the electrons provides a strong constraint, 
$t_{\rm c}\sim3.2\times 10^6 \gamma_3^{-1}$\,yr. 
Radio-emitting electrons with
$\gamma\sim 10^5$, and a plausible galactic field of 1$\mu$\,G, thus
have a lifetime of only $3\times 10^4$\,yr. If these particles diffuse
out (without further acceleration) at a velocity $v\sim0.1\,c$ then they
will have gone only 1\,kpc while emitting in the radio. 
X-ray emitting electrons with $\gamma\sim10^3$ 
can have travelled 100 times further. So the lack of any
corresponding strong radio emission from the X-ray blob could just be
due to Compton losses on the highest energy electrons required to make
that radio emission.

We note that the optical limits may require that the electron
population has a low energy turnoff at $\gamma< 100$.

If the X-ray emission is produced instead within a relativistic jet
with $\Gamma\sim 10$, the required $\gamma$ from the X-rays drops by
10 and the lifetime of the electrons $t_{\rm c}$ 
by a similar factor.
The magnetic field can now be higher without any conflict with the
radio limit. Deep radio and optical imaging will provide valuable
information with which to discriminate between the possibilities.

It is thus likely that the luminous X-ray blob is a by-product of the
powerful inner blazar jet which must have been in operation for 
at least $10^5$\,yr (the light crossing time to the blob).
It only represents about one per cent of
the power of the blazar but is emitted at considerable distance from
the nucleus. There should be energetic protons and nuclei associated
with the electrons which mean that significant energy is being
deposited into the halo of the host galaxy. 
The fact that it is best
seen in the X-ray band is a consequence of the steep $(1+z)^4$
increase in the energy density of the CMB target photons for inverse
Compton scattering. 
The predominance of extended inverse Compton X-ray
emission in distant radio sources is becoming apparent from the results on
3C\,294 at $z$=1.786 (Fabian et al.\ 2003a), 3C9 (Fabian, Celotti \&
Johnstone 2003b) and 4C\,41.17 (Scharf et al.\ 2003). Further high
resolution X-ray observations of distant radio sources should give
more insight into this energy channel, which could influence the
formation and growth of galaxies, groups and clusters.

\section*{Acknowledgements}
ACF thanks the Royal Society for their support.
AC acknowledges the MIUR (COFIN) and ASI for financial support. 
The Chandra data presented in this paper were obtained from the
CXC Chandra archive. We thank all the members of the Chandra team
for making the observation and data analysis available.
This work has made use of the INT-WFC optical imaging data 
obtained by the XID imaging programme of 
the XMM Survey Science Centre (XMMSSC).
This research has made use of the NASA/IPAC Extragalactic Database (NED). 

\bsp

\label{lastpage}

\end{document}